\documentclass[aps,pre,reprint,superscriptaddress,amsmath,showpacs]{revtex4-1}

\usepackage[breaklinks,colorlinks=true]{hyperref}
\usepackage{graphicx}
\usepackage{xcolor}

\begin{document}

\title{Aggregation of self-propelled particles with sensitivity to local order}
\author{Kunal Bhattacharya}
\affiliation {Department of Industrial Engineering and Management, Aalto University School of Science, 00076, Finland}
\affiliation {Department of Computer Science, Aalto University School of Science, 00076, Finland}

\author{Abhijit Chakraborty}
\email[Corresponding author; ]{chakraborty.abhijit.7y@kyoto-u.ac.jp}
\affiliation {Complexity Science Hub Vienna, Josefstaedter Strasse 39, 1080 Vienna, Austria}
\affiliation{Graduate School of Advanced Integrated Studies in Human Survivability, Kyoto University, 1 Nakaadachi-cho, Yoshida, Sakyo-ku, Kyoto 606-8306, Japan}
\date{\today}

\begin{abstract}
We study a system of self-propelled particles (SPPs) in which individual particles are allowed to switch between a fast aligning and a slow nonaligning state depending upon the degree of the alignment in the neighborhood. The switching is modeled using a threshold for the local order parameter. This additional attribute gives rise to a mixed phase, in contrast to the ordered phases found in clean SPP systems. As the threshold is increased from zero, we find the sudden appearance of clusters of nonaligners. Clusters of nonaligners coexist with moving clusters of aligners with continual coalescence and fragmentation. The behavior of the system with respect to the clustering of nonaligners appears to be very different for values of low and high global densities. In the low density regime, for an optimal value of the threshold, the largest cluster of nonaligners grows in size up to a maximum that varies logarithmically with the total number of particles. However, on further increasing the threshold the size decreases. In contrast, for the high density regime, an initial abrupt rise is followed by the appearance of a giant cluster of nonaligners. The latter growth can be characterized as a continuous percolation transition. In addition, we find that the speed differences between aligners and nonaligners is necessary for the segregation of aligners and nonaligners.
\end{abstract}


\maketitle

\section{Introduction}

Collective motion observed in diverse natural and artificial systems has been the subject of numerous experimental and theoretical investigations. Systems that have been studied include fish schools~\cite{huth1994simulation}, birds flocks~\cite{cavagna2010scale}, bacterial colonies~\cite{czirok1996formation}, human crowds~\cite{silverberg2013collective}, as well as, synthetic microswimmer assemblies~\cite{elgeti2015physics} and robotic swarms~\cite{rubenstein2014programmable}. Local interactions in such systems are understood to lead to the emergence of global order or flocking states. This has been demonstrated in self-propelled particle (SPP) models where particles are attributed the tendency to align their direction of motion with their immediate spatial neighbours in the presence of noise~\cite{vicsek1995novel}. Recent studies have also focussed on the possible effects of environmental and individual-level inhomogeneity on the flocking dynamics~\cite{chepizhko2013optimal,copenhagen2016self,yllanes2017many}. For example, disorder is introduced in SPP models in the form of spatially distributed obstacles~\cite{chepizhko2013optimal,chepizhko2013diffusion}, or, a finite fraction of the particles are made non-aligners~\cite{copenhagen2016self,yllanes2017many}. The dynamics in these systems shows the development of phases with complex features like quasi-long-range order~\cite{chepizhko2013optimal} and self-sorting \cite{copenhagen2016self}. In natural flocks, the latter type of inhomogeneity could result from differences in signalling and receptive behaviour or conflict in intentions. In an otherwise homogeneous flock, behavioural shifts at individual levels could imply a certain fraction of flock members spontaneously modifying their nature of motion.       

Flocks of living organisms are known to arrange themselves into cohesive and sometimes segregated units while performing activities like foraging and migration \cite{conradt2005consensus,petit2010decision,bhattacharya2014collective}. This is achieved through consensus decision making by flock members while performing the activities  that, in turn is a consequence of mechanisms at the level of individuals. Such behavioural transitions between different states have been documented in various species \cite{buhl2006disorder,daruka2009phenomenological,miller2012schooling,ginelli2015intermittent}. Suitable modifications to simple SPP models have proven to be useful in reproducing the spatio-temporal features of flocks with behavioural shifts. Models have considered additional attributes to SPPs, like adaptive speed~\cite{buhl2006disorder,li2007adaptive,farrell2012pattern}, random fields~\cite{bhattacharya2010collective}, and transition rates~\cite{ginelli2015intermittent}. 

Experiments on fish schooling~\cite{katz2011inferring,tunstrom2013collective} and bacterial suspensions~\cite{cisneros2011dynamics} have shown that the individual speeds can vary depending on the local order parameter (polarization). A model motivated by these experiments considered SPPs with alignment interactions and a power-law dependence of the speed on the local polarization~\cite{mishra2012collective,singh2020phase}. This showed the nucleation of static clusters and an inverse correlation between the speed and the local density. Notably,  for  SPP  systems  without  alignment such a speed-density  relationship leads to a motility-induced phase separation (MIPS)~\cite{cates2015motility} whereby two phases with distinct densities coexist in the system. This is known to arise from the feedback between the slowing down and crowding of the particles. In general, the variability in the speed both in the absence~\cite{tailleur2008statistical,fily2012athermal,martin2021statistical} and in the presence~\cite{farrell2012pattern,mccandlish2012spontaneous,barre2015motility,sese2018velocity,van2019interrupted} of alignment has been shown to result in novel complex phenomena in models and experiments.

In this paper we study an SPP model where particles can switch between a fast aligning state and a slow non-aligning state depending on the local orientational order parameter. An aligner becomes a non-aligner once the local polarization falls below a threshold $\phi_{th}$, and conversely, a non-aligner becomes an aligner if the local polarization rises above $\phi_{th}$. Using the model we illustrate a mechanism where processing of local information allows an SPP system to simultaneously organize into polarized moving clusters as well as aggregations. The collisions between clusters play a crucial role in such phase separation as we explain later. With our numerical analyses we primarily focus on characterizing the clustering behaviour.


In the absence of a threshold, or equivalently with $\phi_{th}\to 0$, the model expectedly shows a order to disorder transition with the increase in noise~\cite{vicsek1995novel,gregoire2004onset}. We find that the introduction of a finite threshold has a complex interplay with this transition. In the steady state, the model with $\phi_{th}>0$ permits clusters of aligners to coexist with those of non-aligners. The dynamics are found to crucially depend on the level of the noise, the value of $\phi_{th}$ and the overall density. At low noise, the aggregation behaviour of non-aligners can be broadly categorized into two different regimes. For low enough densities, an optimal value of $\phi_{th}$ is found to limit the growth of the largest cluster of non-aligners; and at higher densities the latter is able to grow macroscopically large when $\phi_{th}$ is increased. 

Recent studies have considered SPP models relevant to the understanding of epidemic and information spreading in populations of motile agents~\cite{paoluzzi2018fractal,paoluzzi2020information}. The particles could either irreversibly or reversibly switch between motile and non-motile states, and collectively exhibited fractal aggregation and MIPS. These models considered switching rules based on logic gates involving the states of colliding particles. In contrast, the particles in our model change their states depending upon the orientations of their neighbours.

The outline of the paper is as follows. We explain the details of the model in section~\ref{model}. In section~\ref{results} we discuss the results of our numerical investigation, and in section~\ref{conclusions} we conclude with a summary and final observations.


\section{The Model}
\label{model}
\begin{figure*}
\centering
\includegraphics[width=0.9\textwidth]{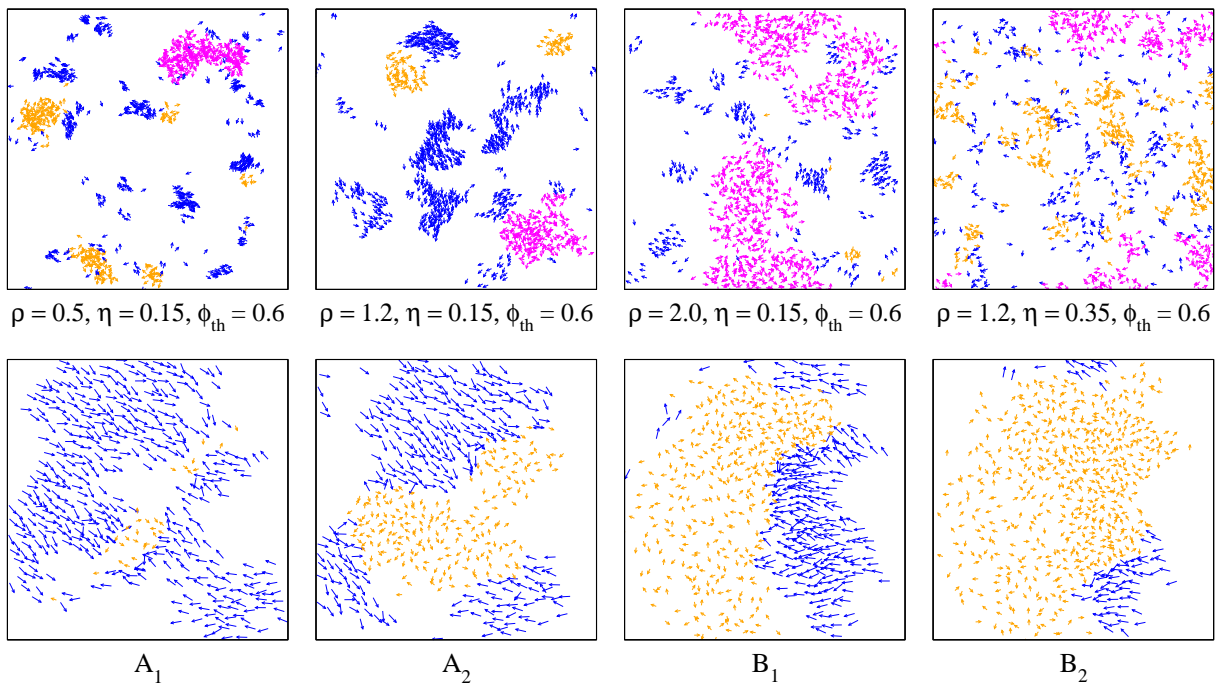}
\caption{Snapshots of the dynamics are shown. The top row corresponds to different parameter sets. The first three (from left) snapshots illustrate the effect of increasing the density ($\rho$), such that, in the third we find an emerging giant cluster of non-aligners. The aligners are denoted in blue (dark gray) and the non-aligners in orange (light gray). The direction of a vector points to the instantaneous direction of motion. Particles belonging to the largest cluster of non-aligners are marked with magenta (gray). In the fourth, the effect of increasing the noise ($\eta$) is shown. In the bottom row, A$_1$ and A$_2$ illustrate the mechanism of the emergence of cluster of non-aligners from the collision of two moving clusters of aligners. The snapshots B$_1$ and B$_2$ show the growth of a cluster of non-aligners when a moving cluster of aligners gets impacted. For both the pairs (A$_1$,A$_2$) and  (B$_1$,B$_2$), the separation in time is $50$ steps which is equivalent to $2.5$ units measured in the time scale $r_0/v_1$. The number of particles in all the snapshots is $N=1024$. For the bottom row, $\rho=1.2$, $\eta=0.15$ and $\phi_{th}=0.6$. Movies corresponding to the first three snapshots (from left) are provided in the Supplemental Material~\cite{sm}.}
\label{fig:fig1}
\end{figure*}

We consider $N$ self-propelled particles to be moving on a two dimensional square area of linear size $L$ under periodic boundary conditions. The global density of the system is given by $\rho=N/L^2$. At discrete times $t$, the state of the $i$th particle is given by its position $\mathbf{r}_i^t$, angle of the direction of motion $\theta_i^t$, and $s_i^t$, denoting aligner ($s_i^t=1$) or non-aligner ($s_i^t=0$). The variables are updated in the following way. First, the  set of neighbours (${\cal N}_i$) of the $i$th particle is enumerated, which comprises of all the particles that are within a distance of $r_0$ from $i$. Then the local order parameter ($\phi_i$), that is the average normalized velocity within the neighbourhood, is calculated as
\begin{equation}
\label{eq:lop}
\phi_i^t=\frac{1}{1+k_i}\lvert\mathbf{n}_i^t+\sum_{j \in \mathcal{N}_i}\mathbf{n}_j^t\rvert.
\end{equation}
Here, $\mathbf{n}_i^t=(\cos\theta_i^{t},\sin\theta_i^{t})$ is a unit vector pointing in the direction of motion of $i$, and $k_i$ is the number of neighbours of $i$. Whether the particle $i$ would have the tendency to align its direction of motion with its neighbours is decided depending on $\phi_i^t$:
\begin{equation}
\label{eq:switch}
   s_i^{t+1}= 
\begin{cases}
    0,& \text{if } \phi_i^t\leq\phi_{th}\\
    1,              & \text{if } \phi_i^t>\phi_{th},
\end{cases}
\end{equation} 
where $\phi_{th}$ is a parameter in the model. Lastly, the angle of heading and the position are updated according to:
\begin{alignat}{2}
\theta_i^{t+1}&=\arg\bigg[ \mathbf{n}_i^t+s_i^{t+1}\sum_{j \in \mathcal{N}_i}\mathbf{n}_j^t+\alpha \sum_{j \in {\cal N}_i} f_{ij}^t\hat{\mathbf{r}}_{ij}^t\bigg]+\eta\xi_i^t,\label{eq:dom}\\
\mathbf{r}_i^{t+1}&=\mathbf{r}_i^{t}+v(s_i^{t+1})\mathbf{n}_i^{t+1},
\label{eq:position}
\end{alignat}
where $v(s_i^{t+1})$ is the magnitude of the velocity depending on whether the particle $i$ is an aligner or a non-aligner, $f_{ij}^t$ is the interaction force between $i$ and its neighbour $j$, $\hat{\mathbf{r}}_{ij}^t$ is a unit vector from $i$ towards $j$, $\alpha$ is the strength of the interaction, $\xi_i^t\in[-\pi,\pi]$ is a delta-correlated angular noise, and $\eta$ is the amplitude of the noise. For an aligner, the dynamics represented in Eqs.~\ref{eq:dom} and~\ref{eq:position} is similar to the Vicsek model~\cite{vicsek1995novel} with an additional short-range interaction between particles \cite{couzin2002collective,gregoire2003moving,gregoire2004onset}. For a non-aligner the second term inside the brackets in Eq.~\ref{eq:dom} is rather absent. 

The force $f_{ij}$ depends on the distance of separation $r_{ij}=|\mathbf{r}_i-\mathbf{r}_j|$, and comprises of finite repulsive and attractive terms:
\begin{equation}
  f_{ij}= 
\begin{cases}
    -f_r\frac{r_e-r_{ij}}{r_e},& \text{if } r_{ij}<r_e\\
\frac{r_{ij}-r_e}{r_a-r_e},& \text{if } r_e<r_{ij}<r_a\\
\frac{r_0-r_{ij}}{r_0-r_a},& \text{if } r_a<r_{ij}<r_0,
\end{cases}
\label{force}
\end{equation}
where $r_e$ is the equilibrium distance, $r_a$ is the distance at which attraction is maximum, and, $f_r$ is the relative magnitude of the repulsive force when $r_{ij}=0$.  The two-body interaction helps to maintain a finite packing density of particles similar to some of the systems~\cite{mccandlish2012spontaneous,van2019interrupted} where MIPS-like phenomena is evidenced. 


In the simulations, we fix the following values for the parameters: $r_0=1$,  $r_a=0.625$, $r_e=0.25$,  $f_r= 1000$, and $\alpha=0.1$. For the non-aligners, we will consider only the repulsive interaction to be present. Most of the results, unless otherwise specified, are obtained with aligner speed, $v_1=0.05$, non-aligner speed, $v_0=0.005$, and noise, $\eta=0.20$. Note, in Eq.~\ref{eq:lop}, $k_i=0$ implies $\phi_i\equiv1$, and hence a non-aligner can switch to the aligning state if it becomes isolated. To ensure that switching arises only as a consequence of interactions in the neighbourhood, we prevent isolated non-aligners from switching states. However, for the parameter ranges that we investigate, this additional rule does not influence the macroscopic behaviour of the system.

\begin{figure*}
\includegraphics[width=0.9\textwidth]{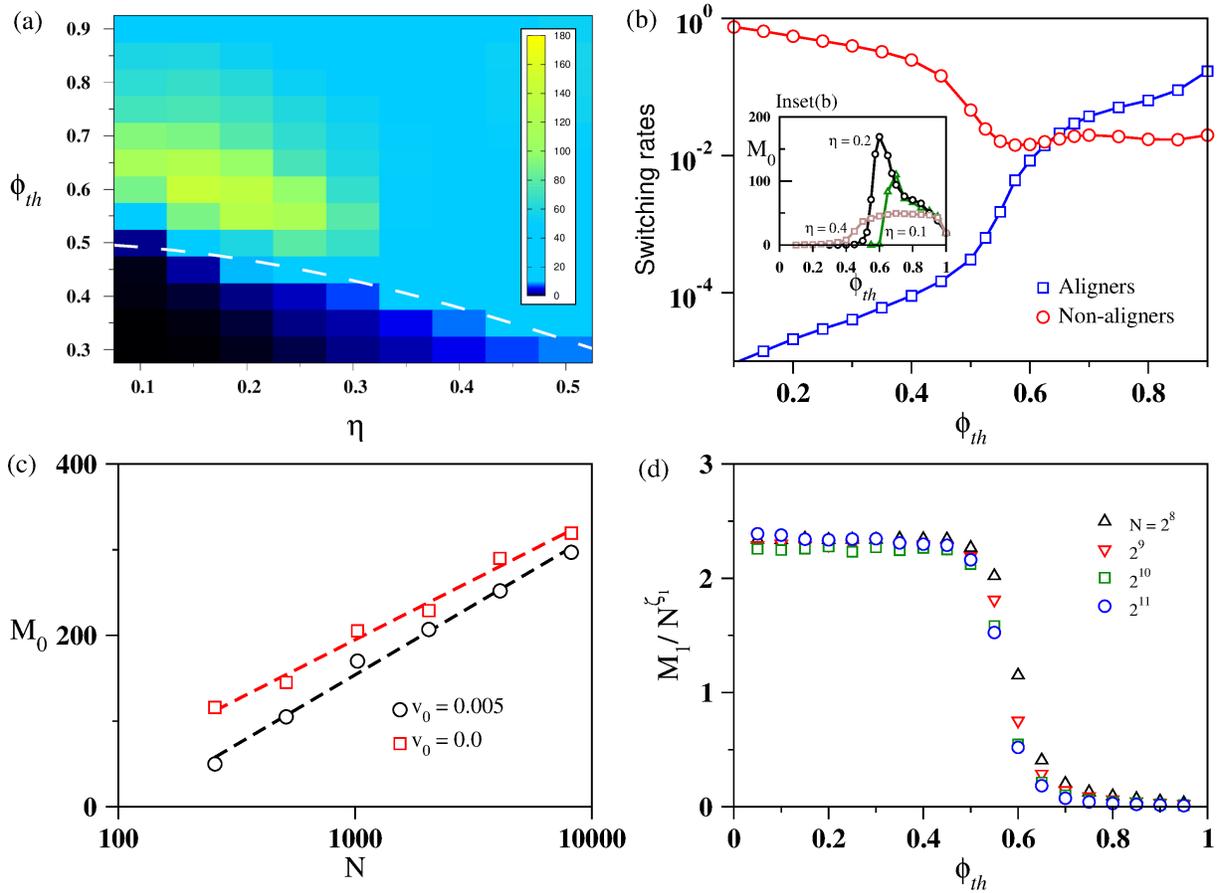}
\caption{Sizes of the largest clusters of non-aligners ($M_0$) and aligners ($M_1$) are characterized at density $\rho=0.5$. 
(a) The dependence of $M_0$ on noise amplitude ($\eta$) and threshold ($\phi_{th}$) is shown as a heat map for $N=2^{10}$. The dashed line represents the equation $\phi_{th}=\phi^{*}(\eta)$ where $\phi^{*}(\eta)$ is given by Eq.~\ref{eq:half-polarization}. (b) The rate of switching per unit time per particle is plotted against $\phi_{th}$ for $N=2^{10}$ at $\eta=0.2$. Here, the aligners are switching to non-aligners, and vice-versa. (Inset b) The variation of $M_0$ with $\phi_{th}$ for three noise amplitudes. The maximum of $M_0$ for $\eta=0.2$ occurs at $\phi_{th}\sim 0.6$ which corresponds to the crossing of the switching rates. (c) The dependence of $M_0$ on $N$ is shown, where $M_0$ is measured at $\phi_{th}=0.6$ and  $\eta=0.2$. In addition to the non-aligner speed $v_0=0.005$, the dependence is also shown for $v_0=0.0$. The dashed lines represent ordinary least squares fits having the form, $M_0=c_0+c_1\log N$. For $v_0=0.005$, $c_0=-225(24)$ and $c_1=60(3)$; for $v_0=0.0$, $c_0=-333(22)$ and $c_1=70(3)$. (d) The dependence of size of the largest cluster of aligners ($M_1$) on $\phi_{th}$ at $\eta=0.2$. The different symbols correspond to different values of $N$, as indicated in the legend. Data collapse is obtained by scaling $M_1$ by the corresponding $N^{\zeta_1}$ with $\zeta_1=0.78$. 
}
\label{fig:fig2}
\end{figure*}

\section{Results}
\label{results}
At low noise and in the absence of a threshold ($\phi_{th}=0$) the system is in a globally ordered state with a single macroscopically large cluster of aligners ($s_i=1$). With the introduction of the switching behaviour ($\phi_{th}>0$), state of the particles become sensitive to fluctuations occurring locally, and  as a result the non-aligners ($s_i=0$) start appearing in the system that are eventually separated from aligners due to the difference in speeds. In the steady state, we find the system to be phase separated into moving clusters of aligners and diffusing clusters of non-aligners. If we observe the system in the very dilute limit, $\rho<0.1$ and with $\phi_{th}=0$, we observe a phase with very small-sized clusters of aligners due to the short-range two-body force~\cite{gregoire2003moving}. This is different from the gaseous phase predicted for the original Vicsek model~\cite{solon2015pattern,chate2020dry}. This also implies that for the higher densities and for finite $\phi_{th}$ the large clusters of aligners have a higher chance to coexist alongside clusters of non-aligners.

Our definition of a cluster is based on connecting neighbouring particles that are in similar states ($r_{ij}<r_0$ and $s_i=s_j$). In Fig.~\ref{fig:fig1} (top row) we show snapshots of steady state configurations resulting from different parameter values. Large fluctuations in $\phi_i^t$ primarily occur as a result of collisions between clusters. Clusters of non-aligners form and grow when moving clusters of aligners collide between themselves or with clusters of non-aligners. This process is illustrated in the bottom row of Fig.~\ref{fig:fig1}. Similarly, when a moving cluster of aligners grazes a cluster of non-aligners, particles at the boundary of the latter switch their states in a short time span to become aligners. Switching of particles at the boundary of a cluster of non-aligners also happen as random events. To describe the generic properties of the system we measure the sizes of the largest clusters of aligners and non-aligners as functions of overall density and speed of the particles. 


\begin{figure*}[t]
\centering
\includegraphics[width=0.9\textwidth]{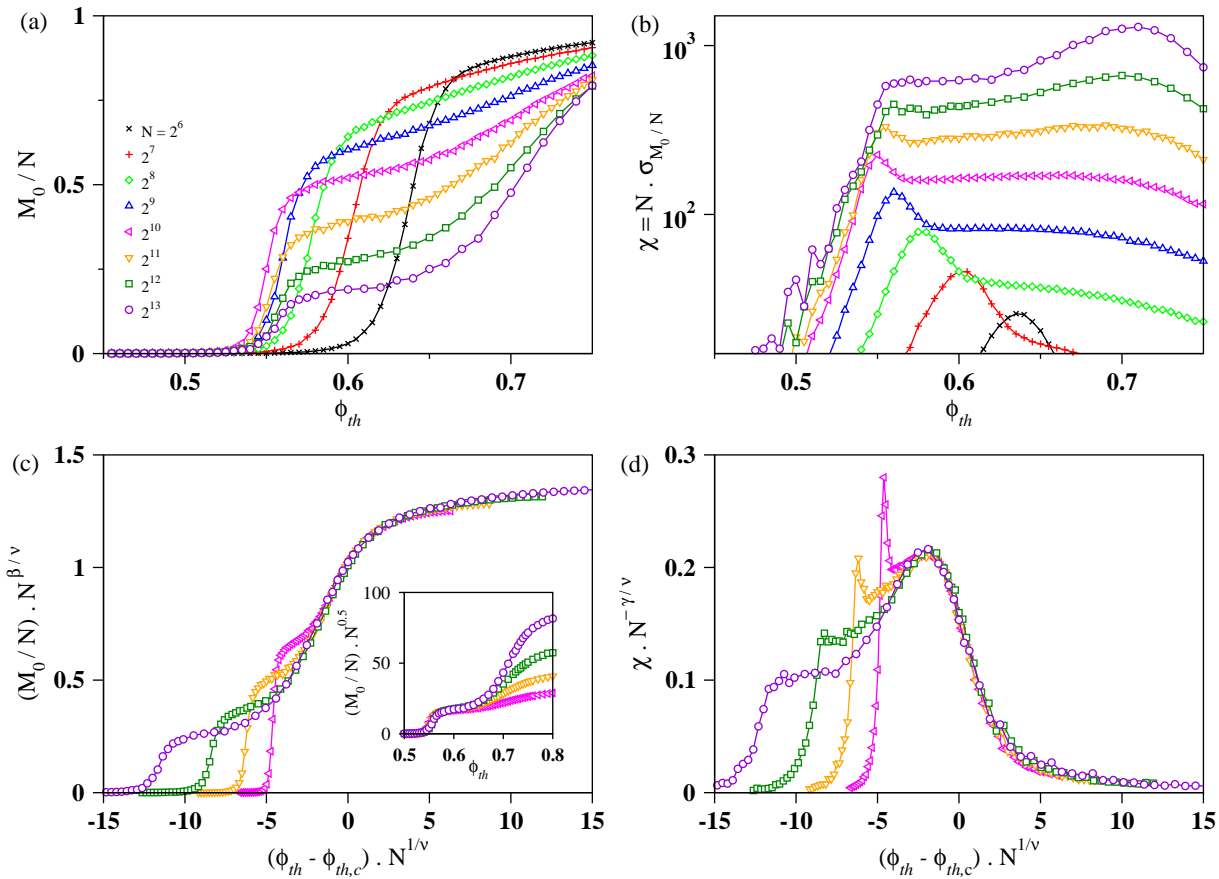}
\caption{(a) The dependence of size of the largest cluster of non-aligners ($M_0$) on the threshold ($\phi_{th}$) at noise, $\eta=0.2$, and density, $\rho=1.6$. The different symbols correspond to different system sizes ($N$), as indicated in the legend. The curves reveal a system size dependent crossover occurring at $N\sim 2^{10}$. For larger systems $M_0$ has an initial steep rise and then a gradual increase.
(b) A susceptibility function $\chi$ corresponding to $M_0$ is shown, where $\chi = N\sqrt{\langle (M_0/N)^2\rangle -  \langle (M_0/N)\rangle^2}$. From around the crossover system size a second peak in $\chi$ starts becoming the dominant maximum. The first peak corresponds to $\phi_{th}=\phi^{*}$ and second peak corresponds to the second increase of $M_0$ at $\phi_{th}=\phi_{th,c}$. (c) Curves from (a) with system sizes from $N\sim 2^{10}$ and above, are rescaled. By plotting $(M_0/N)N^{\beta/\nu}$ versus $(\phi_{th}-\phi_{th,c})N^{1/\nu}$ the validity of Eq.~\ref{OP} is illustrated. Here, $\phi_{th,c}=0.74$. (Inset) Shows $(M_0/N)N^{0.5}$ versus $\phi_{th}$. The collapse shows that the initial rise of $M_0$ at $\phi_{th}=\phi^{*}$ occurs  according to $M_0\sim N^{0.5}$. (d) Collapse of the second peak of susceptibility (from (b)) is obtained following a procedure similar to that in (c). This demonstrates the scaling ansatz in Eq.~\ref{chi}.    
}
\label{fig:fig3}
\end{figure*}

\subsection{Low density regime -- dependence on threshold and noise}

In Fig~\ref{fig:fig2} we show the behaviour of system at a density $\rho=0.5$. The aggregation of non-aligners is only possible when the corresponding clean SPP system is in the ordered phase. A small but finite $\eta$ ensures the presence of clusters of aligners moving in different directions which can collide and allow clusters of non-aligners to nucleate. The latter can not happen when $\eta$ is large and the system is in a gas-like phase where large clusters of aligners are absent. This is evidenced in Fig.~\ref{fig:fig2}(a) where we show the dependence of the size of the largest cluster of non-aligners, $M_0$ on $\eta$ and $\phi_{th}$. The plot also shows that $M_0$ attains its maximum around $\phi_{th}=0.6$ and 
$\eta=0.2$. 

For the individual particles in a cluster of aligners $\phi_i^t$ is high in the ordered phase. However, during the collisions $\phi_i^t$ for particles at the border of the colliding peripheries decreases momentarily. If the drop in  $\phi_i^t$ is less than $\phi_{th}$, then aligners switch to become non-aligners. Therefore, an increase of $\phi_{th}$ leads to an increase of switching events. For aligners we measure the rate of switching as the number of switches to non-aligning states per unit time per aligner. Similarly, we measure the switching rate for non-aligners. As Fig.~\ref{fig:fig2}(b) shows this rate for aligners increases with $\phi_{th}$ and decreases for non-aligners. Initially, at low $\phi_{th}$ the rate is much higher for non-aligners implying non-aligners do not persist and proliferate, but when formed almost instantly switch back to become aligners. But at higher values of $\phi_{th}$ the switching rate for aligners overtakes that for non-aligners. This aspect is also reflected when $M_0$ versus $\phi_{th}$ is examined detail. As shown in Fig.~\ref{fig:fig2} (Inset b) for different $\eta$'s, $M_0$ has a sharp rise at a $\phi_{th}$ and then reaches a maximum. For $\eta=0.2$ the maximum occurs at around $\phi_{th}=0.6$ and coincides with the point where the rates cross each other. The sharp rise in $M_0$ is also found to be noise dependent. Note that the switching rate for the non-aligners becomes  relatively a constant when $\phi_{th}$ is large. This is because inside the bulk of a cluster the average separation is $r_e$ which also implies the number of neighbours for a particle is $k\sim r_0^2/(r_e/2)^2$. With $k$ randomly oriented neighbours inside a cluster of non-aligners the local order parameter takes the typical value of $\phi\sim1/\sqrt{k}\sim r_e/(2r_0)$ in the steady state~\cite{ginelli2016physics}.


An approximation for the
value of the $\phi_{th}$ for which $M_0$ sharply rises can be found in the following way. We consider a non-aligner  on a colliding boundary as illustrated in Fig.~\ref{fig:fig1} (bottom row). We assume that on an average half of the neighbours are non-aligners, and the rest half are aligners. Therefore, the value of the local order parameter can be approximated as
\begin{equation}
    \phi^{*}(\eta)=\frac{1}{2}\frac{\sin \eta\pi}{\eta\pi},
    \label{eq:half-polarization}
\end{equation}
where the expression to the right is half the polar order in an SPP system at low noise and at sufficiently high densities~\cite{peruani2010cluster,dossetti2009phase}. The equation $\phi_{th}=\phi^{*}(\eta)$ is shown as a dashed line in Fig.~\ref{fig:fig2}(a).

In Fig.~\ref{fig:fig2}(c) we find that the maximum values of $M_0$ increases as $\log N$.  In the same plot we show the case of $v_0=0$ where the non-aligners can only rotate but not move. The dependence on $N$  is qualitatively similar in the both the cases. While for $v_0=0$ the ejection of aligners from a cluster of non-aligners occurs due to random switching events, with $v_0>0$ there is an additional diffusion of the non-aligner particles before the switchings happen. In Sec.~\ref{subsec:speed} we show the dependence of $M_0$ on $v_0$.



The abrupt increase in $M_0$ as $\phi_{th}$ increases in the low noise regime coincides with a decrease in the size of the largest cluster of aligners ($M_1$). This is visible in Fig~\ref{fig:fig2}(d) where $M_1$ is plotted as a function of $\phi_{th}$ for different $N$ at $\eta=0.2$. By tuning the exponent in the relation $M_1\sim N^{\zeta_1}$ we obtain the best collapse for different $N$ with $\zeta_1=0.78$.

\begin{figure}
\centering
\includegraphics[width=0.9\columnwidth]{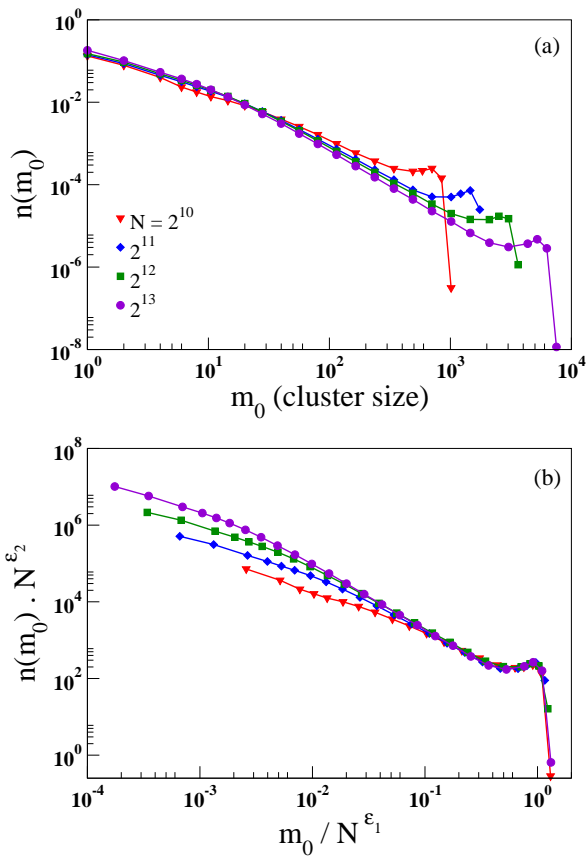}
\caption{
(a) Normalized distributions of cluster sizes of non-aligners for different system sizes, indicated in the legend. The distributions have been obtained at $\eta=0.2$, $\rho=1.6$ and $\phi_{th}=0.74=\phi_{th,c}$. (b) A collapse of the distributions from (a), obtained by plotting $n(m_0)N^{\epsilon2}$ versus $m_0/N^{\epsilon_1}$ and tuning 
$\epsilon_1$ and $\epsilon_2$. In (b) $\epsilon_1=0.96(2)$ and $\epsilon_2=1.98(1)$. The collapse indicates a power-law $n(m_0)\sim {m_0}^{-\tau}$ in the large $N$ limit with $\tau=\epsilon_2/\epsilon_1=2.06(3).$
}
\label{fig:fig4}
\end{figure}

\subsection{High density regime}
\label{subsection:hd}

In the high density and low noise regime, we observe the largest cluster of non-aligners grows with $\phi_{th}$ whose size is of the order of $N$ when $\phi_{th}$ is unity. An incipient cluster in this regime is shown in Fig.~\ref{fig:fig1} (top row - third from left). In addition, we find the behaviour of the system to be strongly dependent on $N$.  In Fig.~\ref{fig:fig3}(a) we plot the fraction $M_0/N$ for different system sizes. From the different curves we observe that the generic dependence of $M_0/N$ on $\phi_{th}$ is different in smaller and larger values of $N$ with a crossover occurring at $N\sim2^{10}$. For the smaller systems the growth in the largest cluster primarily occurs when $\phi_{th}$ is between 0.6 and 0.7. For the larger systems there is an initial rapid increase in $M_0$ that is similar to that observed at low densities, and is followed by a gradual further growth.

Considering the growth of the larger cluster of non-aligners to be a percolation phenomenon that occurs with respect to the tuning of $\phi_{th}$, we calculate a susceptibility corresponding to the order parameter $M_0/N$, given by $\chi = N \sigma$, with $\sigma^2= \langle (M_0/N)^2 \rangle -  \langle (M_0/N) \rangle^2$~\cite{radicchi2009explosive}, where angular brackets denote averaging in the steady state. The plot of $\chi$ as a function of $\phi_{th}$ in Fig.~\ref{fig:fig3}(b) demonstrates a crossover in the finite-size effect. For $N<2^{10}$ there is only a single maximum that shifts to the left of the $\phi_{th}$ axis on increasing $N$. With $N\geq 2^{10}$ we find the emergence of a second peak in $\chi$ which is the dominant maximum as $N$ increases further. 
Ideally, for a given $N$, the position of the (second) maximum of $\chi$ is expected to provide the critical thresholds (pseudocritical point) $\phi_{th,c}(N)$. Observing that accurately locating the second maximum can be difficult for the smaller system sizes we circumvent the problem in the following way. For the finite-size effects in percolation we assume the relations,

\begin{eqnarray}
    M_0/N&=&N^{-\beta/\nu}\mathcal{F}[(\phi_{th}-\phi_{th,c})N^{1/\nu}],\label{OP}\\
    \chi&=&N^{\gamma/\nu}\mathcal{G}[(\phi_{th}-\phi_{th,c})N^{1/\nu}],\label{chi}
\end{eqnarray}
where $\phi_{th,c}$ is the critical threshold in the infinite size limit ($N\to \infty$), and $\nu$, $\beta$, and $\gamma$ are the critical exponents characterizing a second order percolation transition. Using the above two
relations and the fact that $\chi=N\sigma$ we get a hyperscaling relation,

\begin{equation}
    \gamma/\nu = 1-\beta/\nu.
    \label{hyperscaling}
\end{equation}

At different values of $\phi_{th}$ we fit power laws to the data corresponding to $M_0/N$ versus $N$. This gives us a set of trial values for the exponent $\beta/\nu$. Similarly, we obtain a set of trial values of $\gamma/\nu$ from $\chi$ versus $N$ at different $\phi_{th}$. Then we obtain the critical point $\phi_{th,c}$ by locating the $\phi_{th}$ at which $\beta/\nu$ and $\gamma/\nu$ satisfies Eq.~\ref{hyperscaling}. This method yields $\beta/\nu=0.035(9)$, $\gamma/\nu=0.965(4)$, and $\phi_{th,c}=0.740(5)$. The error estimates in the exponents correspond to power law fits at $\pm0.005$ from $\phi_{th,c}$. (In the Appendix we provide expressions for $M_0$ and $\phi_{th,c}$ from a reaction-limited description.)

To determine $\nu$, we first scale the y-axis of Fig.~\ref{fig:fig3}(a) by multiplying with $N^{\beta/\nu}$. Then upon fixing a value of $(M_0/N)N^{\beta/\nu}$ around $1.0$ we obtain the corresponding values of $\phi_{th}$ for different $N$. We estimate the value of $1/\nu$ from the slope of the line fitted with $\log\vert \phi_{th} \left(N\right) -\phi_{th,c} \vert$ versus $\log N$. Repeating the process for different values of $(M_0/N)N^{\beta/\nu}$, we get $\nu=2.16(3)$. Using the above values for the scaling exponents and $\phi_{th,c}$ we obtain data collapse for $M_0/N$ and $\chi$ shown in Fig.~\ref{fig:fig3}(c) and Fig.~\ref{fig:fig3}(d), respectively. The collapse of the curves for different $N$ when $\phi_{th}$ is close to $\phi_{th,c}$ shows that the scaling forms in Eq.~\ref{OP} and~\ref{chi} hold true. 

Similar to the case of low density, there is an initial rapid increase in $M_0$ at $\phi_{th}=\phi^{*}$. This is characterized by solely scaling the $M_0/N$ axis,  and collapsing the curves for different $N$ as shown in the inset of Fig.~\ref{fig:fig3}(c).  The scaling shows that as $\phi_{th}$ crosses $\phi^{*}$, $M_0$ abruptly increases from $O(1)$ to $O(\sqrt{N})$. The latter increase in $M_0$ also coincides with a fall in $M_1$ (not shown). For $\phi_{th}<\phi^{*}$ we find $M_1\sim N^{\zeta_2}$ with $\zeta_2=0.80$ that is quite close to $\zeta_1$.

Additionally, we also studied the cluster size distribution of the non-aligners. We obtained the statistics by observing the systems at $\phi_{th} = \phi_{th,c}$. The normalized distributions $n(m_0)$ for the cluster sizes ($m_0$) of non-aligners are plotted in Fig.~\ref{fig:fig4}(a) for different $N$. We assume power-law distributed cluster sizes and finite-size effects are present, such that  $n(m_0)\sim m_0^{-\tau}f\left(m_0/M_0\right)$, where $\tau$ is the Fisher exponent. The system being at $\phi_{th,c}$ we expect $M_0\sim N^{\epsilon_1}$, with $\epsilon_1=1-\beta/\nu$, and therefore, the above power-law can be recast into the following form:
\begin{equation}
    n(m_0)=N^{-\epsilon_2} D\left( m_0/N^{\epsilon_1} \right),
    \label{eq:power_law}
\end{equation}
where, the scaling function $D(x)\sim x^{-\tau}$ for $x\to 0$, and  $D(x)$ decreases faster than a power law for $x\gg 1$. This implies $\tau=\epsilon_2/\epsilon_1$. In Fig.~\ref{fig:fig4}(b) we plotted $n(m_0)N^{\epsilon_2}$ versus $m_0/N^{\epsilon_1}$ and tuned the values of $\epsilon_1$ and $\epsilon_2$ to get a collapse of the distribution for different $N$. The latter allows us to validate Eq.~\ref{eq:power_law}. We get the best collapse for $\epsilon_1=0.96(2)$ and $\epsilon_2=1.98(1)$, which implies $\tau=2.06(3)$.

\begin{figure}
\centering
\includegraphics[width=0.9\columnwidth]{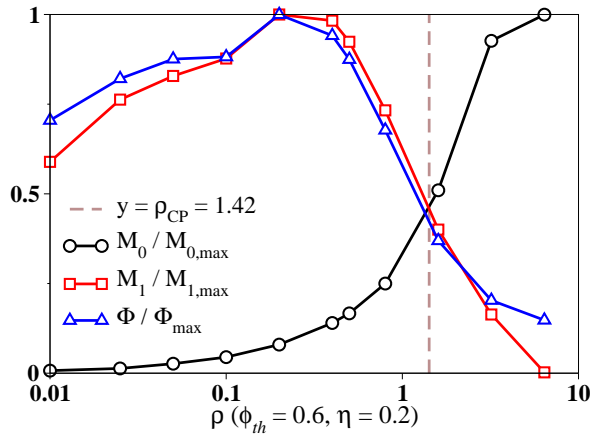}
\caption{The following quantities of interest are plotted as functions of density ($\rho$) at noise, $\eta=0.2$, and threshold, $\phi_{th}=0.6$: $M_0/M_{0,\textrm{max}}$ -- size of the largest cluster of non-aligners normalized by its maximum value within the range of investigation,  $M_1/M_{1,\textrm{max}}$ -- normalized size of the largest cluster of aligners, and, $\Phi/\Phi_{\textrm{max}}$ -- normalized polarization for the system. The dashed vertical line indicates the density $\rho_{CP}=1.42$ that corresponds to the critical filling factor for continuum percolation. }
\label{fig:fig5}
\end{figure}

\begin{figure}
\centering
\includegraphics[width=0.9\columnwidth]{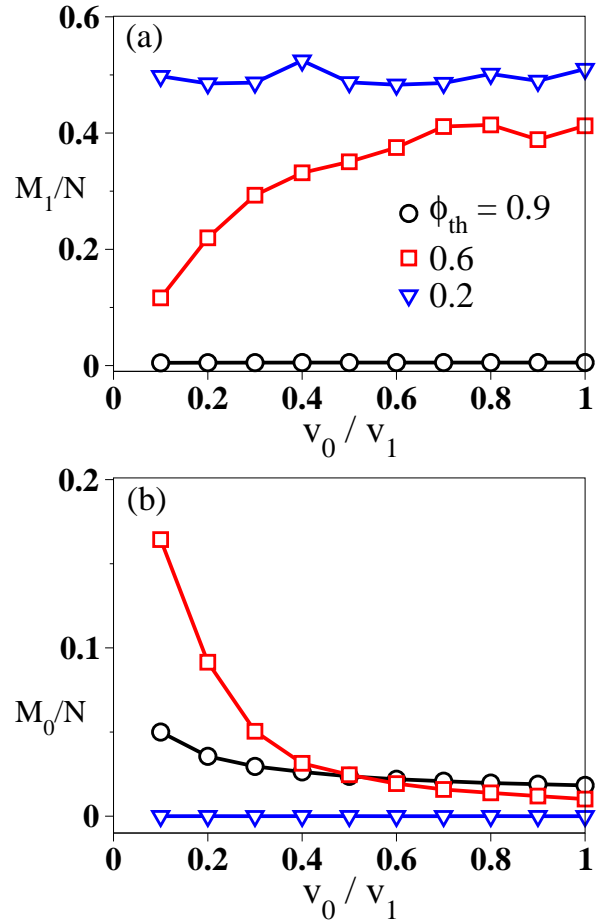}
\caption{(a) Dependence of size of the largest cluster of aligners ($M_1$) on non-aligners speed ($v_0$), for three different thresholds ($\phi_{th}$). The aligner speed, $v_1=0.05$, noise, $\eta=0.2$, number of particles, $N=1024$, and density $\rho=0.5$. (b) Dependence of size of the largest cluster of non-aligners ($M_0$).}
\label{fig:fig6}
\end{figure}

\subsection{Generic dependence on density}
After observing that switching of particle states in ordered flocks produces two distinct type of mixed phases -- non-percolating and percolating depending on the density $\rho$, we investigate how the relevant quantities continuously vary as functions of $\rho$. In Fig.~\ref{fig:fig5} we show the variation of $M_0$ and $M_1$ at low noise ($\eta=0.2$) and substantial presence of switching activity ($\phi_{th}=0.6$). In addition to the sizes of largest clusters we also show the polarization $\Phi$ ~\cite{cavagna2010scale, copenhagen2016self} which is a measure of the degree of global order in the system, and is defined as $\Phi = \langle\lvert (1/N) \sum_{i=1,N} \mathbf{n}_i^t\rvert\rangle$. 

We compare $M_0$, $M_1$ and $\Phi$ scaled by their respective maximum values with $\rho$ being in the range $0.01$ to $10.0$. It is known that in a pure SPP system for a fixed $\eta$, long-range order vanishes when density is lowered~\cite{czirok2000collective}. This phenomena partially underlies the dynamics observed in our case where switching is allowed. The relatively low degree of order as reflected in the values of $M_1/M_{1,max}$ and  $\Phi/\Phi_{max}$ at very low densities is result of the lack of order in the pure system. The maximum in the values of $M_1$ and $\Phi$ occur at around $\rho\gtrsim 0.1$. In this regime, the absence of switching implies that the system has high degree of order and a macroscopically large cluster of aligners always exists in the steady state. When switching is present we observe the non-percolating mixed phase where $M_0\sim \log N$. A further increase in the density would result in the enhancement of order and aligner cluster size in the pure system. However, for the mixed phase the aligner cluster size and order decreases with further increase in density. The cluster size for non-aligners increases monotonically. The possibility of the formation of a giant cluster of non-aligners may be considered similar to the situation of continuum percolation (CP) of overlapping disks with radii $R=r_0/2$. Noting that the filling fraction is defined as $\pi R^2\rho$ and that the estimate for the critical value is around $1.12$~\cite{mertens2012continuum}, the corresponding critical density is $\rho_{\textrm{CP}}=1.42$. This density, therefore, would signify the bordering between non-percolating and percolating mixed phases. The latter is evident in the case of $\rho=1.6 > \rho_{\textrm{CP}}$.

\subsection{Dependence on speed differences} 
\label{subsec:speed}
Lastly, we study how the difference between the speeds of the aligners and non-aligners governs the evolution of system. We fix the aligner speeds to $v_1=0.05$ and vary the speed of the non-aligners, $v_0$. We plot $M_1$ and $M_0$ as functions of $v_0$ for three different values of $\phi_{th}$ in Fig.~\ref{fig:fig6}.  It is apparent that the mixed phase ($\phi_{th}$=0.6) where macroscopically large aligner and non-aligner clusters coexist is delicately dependent on the value $v_0$. As $v_0$ approaches $v_1$, $M_0$ is found to decrease and $M_1$ is found to increase. The diffusion of non-aligners occurs with a diffusion constant that is proportional to $v_0^2$~\cite{peruani2010cluster,peruani2013kinetic}. Therefore, the rate of ejection of aligners from the boundary of a cluster of non-aligners also increases as the non-aligner speed is increased. In addition, as the relative difference of speeds vanishes, the non-aligners formed after a collision effectively fail to segregate and to eventually proliferate. As a result of the above the effect of a finite $\phi_{th}$ diminishes and large clusters of non-aligners are rarely observed. These results show that the formation of mixed phases that is controlled by $\phi_{th}$ is also dependent on the difference in speeds. We have shown the dependence on $v_0$ in the non-percolating regime, but the indications are similar for the percolating case as well.

\section{Conclusions}
\label{conclusions}
We studied a system where self-propelled particles were allowed to switch states between fast aligners and slow non-aligners based on the degree of alignment in their neighbourhood. In the steady state, the system segregated into separate clusters of aligners and non-aligners. In the mixed phase, the largest cluster of aligners was found to vary algebraically with the system size. However, depending on the density of the system, the aggregation of the non-aligners appeared to be very different. For the low densities, the largest cluster of non-aligners reached a maximum size for an optimal noise and an optimal threshold. For high densities, after an initial abrupt increase, a giant percolating cluster could emerge with the increase in the threshold. Also, the behaviour for small system sizes appeared to be very different. The boundary between the density regimes roughly coincided with the density corresponding to the critical filling factor for a continuum percolation transition.  Irrespective of the density, the separation of speeds seemed to be a necessary condition for the model to display the segregation of non-aligners. When the speeds become comparable, large clusters of non-aligners are predominantly absent.  Although the appearance of the giant cluster conforms to a set of finite-size scaling hypotheses, the transition could be  non-universal~\cite{liu1997computer} with dependence on the noise amplitude and density, through the functions $\mathcal{F}$ and $\mathcal{G}$ in Eqs.~\ref{OP} and~\ref{chi}. 

 The percolation of clusters was recently studied~\cite{kyriakopoulos2019clustering} in the classical Vicsek model. Unlike our model, the SPPs in the Vicsek model  always remain aligners (without switching) and attraction-repulsion forces are absent. The authors investigated the global connectivity of clusters with increase in the global density ($\rho$) along both the longitudinal and the transverse directions with respect to the direction of global order. They estimated a critical density, $\rho_c=1.96~(>\rho_{CP})$. Similar to the current model, if we denote the size of the largest cluster (of aligners) in the Vicsek model as $M_1$, then near to $\rho_c$, the dependence on $N$ may be characterized by using $M_1\sim N^{\zeta}$. Using the reported~\cite{kyriakopoulos2019clustering} values of the different critical exponents,  $\zeta$ is found be in the range $0.95-1.00$. In our model, for the density regimes investigated, and when the clusters of aligners are macroscopically large ($\phi_{th}< \phi^{*}$)  we find $\zeta$ to be in the range $0.78-0.80$. Taken together, we expect that our model was investigated at densities which are still lower than the critical density that would be needed for the percolation of clusters of aligners if switching is absent. Also note, that in our model, clusters of non-aligners are formed mainly due to the collisions between clusters and the speed difference between aligners and non-aligners.

Recent advances in living active matter have found that modifications in individual behaviour through the sensing of local densities lead to the formation of regions of orientational disorder and aggregations~\cite{ling2019behavioural,leggett2019motility,klamser2021impact,bauerle2018self}. Similar observations are made in experiments with active colloidal systems employing different methods to program the particle motion, like optical feedback loops and field modulations~\cite{bauerle2018self,lavergne2019group,soma2020phase}. Therefore, the observed macroscopic behaviour in our model could be of relevance, for example, to active colloids with setups allowing the particles to sense and respond to the average orientation of neighbours~\cite{bauerle2020formation}, to the design and control of robot swarms~\cite{werfel2014designing,rubenstein2014programmable}, and in general, to systems exhibiting both polar order and MIPS-related behaviour~\cite{van2019interrupted,crosato2019irreversibility,sese2021phase}. Also, owing to the additional state variable in our model, the latter can be contrasted with the study of clustering and percolation in the classical Vicsek model~\cite{kyriakopoulos2019clustering}, and similarities with models on information spreading in motile collectives can be further explored~\cite{paoluzzi2018fractal,paoluzzi2020information}.


Also, there appears to be scope for additional complexity in the current model. We have assumed that switching behaviour is symmetric in terms of having a single threshold for both aligners as well as non-aligners. In a more general scheme, there can be two different thresholds for particles of either types. Currently, the additive noise causes the moving clusters to collide which generates the non-aligners. It would be interesting to test the model at zero-noise~\cite{chakraborty2016spontaneous} but with other forms of disorder like boundaries, obstacles~\cite{chepizhko2013optimal} and quenched non-aligners~\cite{yllanes2017many}. The phase separation between aligners and non-aligners could be also studied in systems without self-propulsion, for example, in Brownian walkers with a velocity alignment interaction~\cite{dossetti2015emergence} and in the Vicsek model on the lattice~\cite{bhattacherjee2014cyclic}.

\section{Appendix}
\label{appendix}
\subsection{Mean field description}
\label{MFT}
Here we provide a reaction-limited description of the system in the high density regime based on our simulations and neglecting the spatial correlations. We consider the system to be consisting of the following types of particles: particles that are part of the giant cluster of non-aligners ($M_0$), aligners ($N_1$), and non-aligners that are not part of the giant cluster ($N_0'$), such that
\begin{equation}
    M_0+N_0'+N_1=N.
    \label{eq:conservation}
\end{equation}

As seen above with regards to Fig.~\ref{fig:fig3}, as $\phi_{th}$ crosses $\phi^{*}$, $M_0$ becomes $O(\sqrt{N})$. On further increasing $\phi_{th}$, when the latter reaches $\phi_{th,c}$, the incipient giant cluster is observed. We model the growth of this cluster using the following equation:
\begin{equation}
\frac{\mathrm{d}M_0}{\mathrm{d}t}=(\phi_{th}-\phi^{*}) AM_0N_1-(1-\phi_{th})BM_0.
\label{eq:m0}
\end{equation}
The first term to the right accounts for the collision between the aligner particles and the giant cluster of non-aligners by which particles are added to the perimeter of the latter. Given that it is a percolating cluster which is growing, and is far from being circular in shape the perimeter is assumed to vary as $M_0$. A typical such cluster is shown in the Fig.~\ref{fig:fig1} (top row - third from the left). In general, for a percolating cluster the relation between the perimeter (or hull) and the mass is given by $H_0\sim {M_0}^x$, with $x=d_{h}/d_f$. Here $d_f$ and $d_h$ are the fractal and hull dimensions, respectively, and can be computed using the relations $d_f=d-\beta/\nu_L$ and $d_h=1+1/\nu_L$~\cite{saleur1987exact,stauffer2003introduction}. In the current context, the exponents from our numerical calculations in Sec.~\ref{subsection:hd} would indicate $x \approx1$. The exponent  $\nu_L=\nu/d$, where the dimensionality of space, $d=2$. For classical percolation in $d=2$, $d_f=91/48$ and $d_h=7/4$, which gives $x\simeq 0.92$~\cite{stauffer2003introduction}. On the other, clusters of aligners are mostly small in size in this regime, therefore, the dependence is taken to be proportional to the total number of aligners, $N_1$. The colliding aligners are expected to have their local order parameter distributed around $\phi^{*}$. Keeping other factors unchanged, as $\phi_{th}$ approaches $\phi^{*}$ from below and eventually crosses $\phi^{*}$, more and more aligners are expected to switch their states.  
We approximate this dependence on $\phi_{th}$ as, $\phi_{th}-\phi^{*}$. The coefficient, $A$ contains factors in the collision rate including the speed of the group of aligners~\cite{peruani2010cluster,peruani2013kinetic} and is independent of $\phi_{th}$. The second term accounts for the loss of particles by which particles near the perimeter of the giant cluster switch to aligning states and detach from the latter.  While within the bulk, the local order parameter would typically be around $r_e/(2r_0)$ (discussion on Fig.~\ref{fig:fig2}(b)), for the non-aligners at the boundary we expect larger fluctuations and hence a flatter distribution extending up to unity. We assume that the rate of switching is proportional to $1-\phi_{th}$. The coefficient $B$ takes into account other factors independent of $\phi_{th}$.

We observe in our simulations that apart from the ones in the giant cluster, the non-aligners are formed in the process of collision between small clusters of non-aligners. The resulting clusters of non-aligners are also small and are not stable. We model this process using the following equation:
\begin{equation}
\frac{\mathrm{d}N_0'}{\mathrm{d}t}=(\phi_{th}-\phi^{*})kA{N_1}^2-(1-\phi_{th})BN_0'.
\label{eq:n0'}
\end{equation}
The first term accounts for formation of the small clusters of non-aligners. We assume that the coefficient in the collision rate accounting for factors independent of time and $\phi_{th}$ is only different by a multiplicative constant from the coefficient in Eq.~\ref{eq:m0}. The second term is similar to the loss term in $\mathrm{d}{M_0}/\mathrm{d}t$.

\begin{figure}
\centering
\includegraphics[width=0.9\columnwidth]{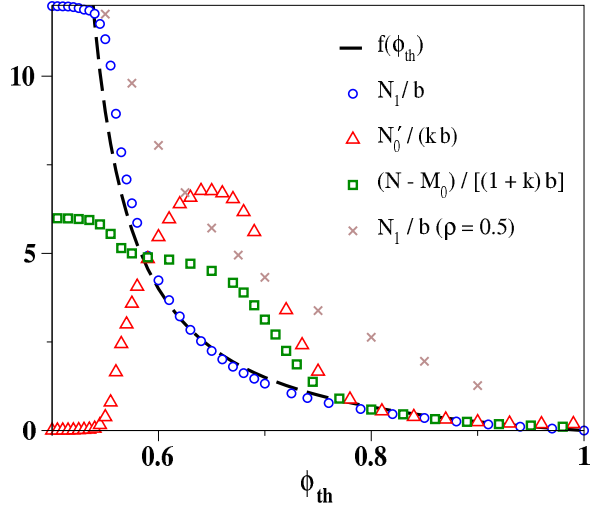}
\caption{Results from the simulation in the high density regime ($\rho=1.6$, $\eta=0.2$, and $N=2^{13}$) is compared with the steady state solutions from the model. The dashed line denotes the function $f(\phi_{th})=(1-\phi_{th})/(\phi_{th}-\phi^{*})$. The circles show the dependence on $\phi_{th}$ for the number of aligners ($N_1$) scaled by a constant ($b$). Choosing $b/N\approx 1/12$ shows the validity of Eq.~\ref{eq:N1s}. Similarly, the triangles and the squares show the ranges of validity for the relation $N_0'=kN_1$ and Eq.~\ref{eq:M0}, respectively where $k=1$. The plot of $N_1/b$ in the low density regime is shown using the crosses. Here, we take $\phi^{*}=0.5$ which provides the best fit instead of $\phi^{*}=0.47$ which is obtained from Eq.~\ref{eq:half-polarization}.}
\label{fig:fig7}
\end{figure}

We are interested in the steady state dependence of $M_0$ on $\phi_{th}$. Therefore, we set the right hand side of Eq.~\ref{eq:m0} to zero. Assuming a non-zero finite solution for $M_0$ we get the following steady solution for $N_1$:
\begin{equation}
N_1^s=bf(\phi_{th}),
\label{eq:N1s}
\end{equation}
where $f(\phi_{th})=(1-\phi_{th})/(\phi_{th}-\phi^{*})$ and $b=B/A$. Similarly, by equating the right hand side of Eq.~\ref{eq:n0'} to zero, we get $N_0'^s=kN_1^s$. 
Next, the steady state solution for $M_0$ is found by using $N_1^s$ and $N_0'^s$ in Eq.~\ref{eq:conservation}:
\begin{equation}
M_0^s=N-b(1+k)f(\phi_{th}).
\label{eq:M0}
\end{equation}
 In Fig.~\ref{fig:fig7} we compare the steady state solutions for $N_1$, $N'_0$ and $M_0$ with the numerical results. As mentioned above, the derived expression appears to be valid for $\phi_{th}>\phi_{th,c}$ when the incipient giant cluster is already present. Below $\phi_{th,c}$, clusters of non-aligners keep continually forming and fragmenting, clusters of aligners are relatively larger in size, and spatial correlations cannot be neglected in the description of the dynamics.
 
 Considering that the collision coefficient $A$ is inversely proportional to the area ($L^2$)~\cite{peruani2010cluster,peruani2013kinetic} we can also rewrite Eq.~\ref{eq:M0} as 
\begin{equation}
 M_0^s=N\left[1-(c/\rho) f(\phi_{th})\right],    
 \label{eq:M0-rho}
\end{equation}
where $c/\rho=b(1+k)/N$. The above expressions for $M_0^s$ imply that when $N$ increases at a fixed density or when $\rho$ increases, the transition to a global connectivity becomes faster. This expression, however, is not valid in the limit $\rho\to 0$.  For the latter limit in Eq.~\ref{eq:m0}, the second term would be the dominant term for all values of $\phi_{th}$ and $M_0$ would be zero in the steady state. Alternately, the dynamics at low density cannot be described by Eqs.~\ref{eq:m0} and~\ref{eq:n0'}, and as a result, Eq.~\ref{eq:N1s} does not hold. The deviation of $N_1^s/b$ at low density from the function $f(\phi_{th})$  is shown in Fig.~\ref{fig:fig7}. 

We obtain an estimate of $\phi_{th,c}$ by assuming that once the critical threshold is exceeded, the number of non-aligners inside the largest cluster becomes greater than the number of non-aligners present outside: $M_0^s>N_0'^s$. Here using the expressions $M_0^s$ and $N_0'^s$, we get a lower bound on $\phi_{th}$, 
\begin{equation}
    \phi_{th,c}=\phi^{*}+\frac{3c}{3c+2\rho}(1-\phi^{*}),
\end{equation}
where we have set $k=1$ as shown in Fig.~\ref{fig:fig7}. Therefore, in this reaction dominated description, the limits $N\to \infty$ at a fixed density or $\rho\to \infty$ imply that $\phi_{th,c}\to \phi^{*}$ from above.  Finally, given that the function $f(\phi_{th})$ is analytic in the range $\phi^{*}<\phi_{th}\leq 1$, we can express as a first order approximation: $M_0^s(\phi_{th})- M_0^s(\phi_{th,c})\sim (\phi_{th}-\phi_{th,c})$ for $\phi_{th}\gtrsim\phi_{th,c}$, giving $\beta{\textrm{(mean-field)}}=1.$

\begin{acknowledgments}
 KB and AC acknowledge helpful discussions with Shakti N. Menon, Sasidevan Vijayakumar, and Sumanta Kundu.
\end{acknowledgments}

%

\end{document}